\documentclass[final,3p,times,twocolumn]{elsarticle}

\usepackage{graphicx}

\usepackage{amsmath}
\usepackage{cases}
\usepackage{upgreek}
\biboptions{sort&compress}

\usepackage{color,soul}
\sethlcolor{yellow}

\sethlcolor{white}

 \usepackage{lineno}

\newcommand*\patchAmsMathEnvironmentForLineno[1]{%
  \expandafter\let\csname old#1\expandafter\endcsname\csname #1\endcsname
  \expandafter\let\csname oldend#1\expandafter\endcsname\csname end#1\endcsname
  \renewenvironment{#1}%
     {\linenomath\csname old#1\endcsname}%
     {\csname oldend#1\endcsname\endlinenomath}}%
\newcommand*\patchBothAmsMathEnvironmentsForLineno[1]{%
  \patchAmsMathEnvironmentForLineno{#1}%
  \patchAmsMathEnvironmentForLineno{#1*}}%
\AtBeginDocument{%
\patchBothAmsMathEnvironmentsForLineno{equation}%
\patchBothAmsMathEnvironmentsForLineno{align}%
\patchBothAmsMathEnvironmentsForLineno{flalign}%
\patchBothAmsMathEnvironmentsForLineno{alignat}%
\patchBothAmsMathEnvironmentsForLineno{gather}%
\patchBothAmsMathEnvironmentsForLineno{multline}%
\patchAmsMathEnvironmentForLineno{cases}%
\patchAmsMathEnvironmentForLineno{numcases}%
}

\journal{Thermochimica Acta}

\begin{document}

\begin{frontmatter}



\title{Absolute accuracy in membrane-based ac nanocalorimetry}


\author{S.~Tagliati and A.~Rydh\fnref{email}}
\fntext[email]{Corresponding author. Tel. +46 8 5537 8692.\\
E-mail address: andreas.rydh@fysik.su.se (A. Rydh).}

\address{Department of Physics, Stockholm University, AlbaNova University Center, SE~--~106~91 Stockholm, Sweden}

\begin{abstract}
To achieve accurate results in nanocalorimetry a detailed analysis and understanding of the behavior of the calorimetric system is required. There are especially two system-related aspects that should be taken in consideration: the properties of the empty cell and the effect of the thermal link between sample and cell. Here we study these two aspects for a membrane-based system where heater and thermometer are both in good contact with each other and the center of the membrane. Practical, analytical expressions for describing the frequency dependence of heat capacity, thermal conductance, and temperature oscillation of the system are formulated and compared with measurements and numerical simulations. We finally discuss the experimental conditions for an optimal working frequency, where high resolution and good absolute accuracy are combined.
\end{abstract}

\begin{keyword}
Nanocalorimetry \sep ac-calorimetry \sep Membrane \sep Absolute accuracy \sep Frequency dependence.


\end{keyword}

\end{frontmatter}


\section{Introduction}
\label{Introduction}
\hl{Thermodynamic measurements with both good absolute accuracy and high resolution are essential to understand fundamental properties of materials.} Demand for nanocaloric measurements is coming both from the wish to study new physics at mesoscopic scales and the need to investigate bulk behavior such as anisotropy and magnetic field dependence of novel materials that are difficult to synthesize as large crystals. A calorimetric method particularly suitable for studying sub-$\upmu$g samples is the ac steady state method \cite{Sullivan,RiouRSI,Minakov,Huth,RydhEMSAT,Garden,TagliatiCalori,Kohama}. This method usually has high resolution but rather low absolute accuracy because of the difficulties related to the choice of the correct working frequency \cite{RiouSM}. A poorly selected working frequency results in thermal disconnection of the sample, or, for small samples, uncontrolled contribution of the frequency dependent \hl{addenda} heat capacity. A detailed analysis of the behavior of the calorimetric system is thus required to know the conditions at which the most accurate results are obtained. 

\begin{figure}[t]
\begin{center}
\includegraphics[width=0.8\linewidth]{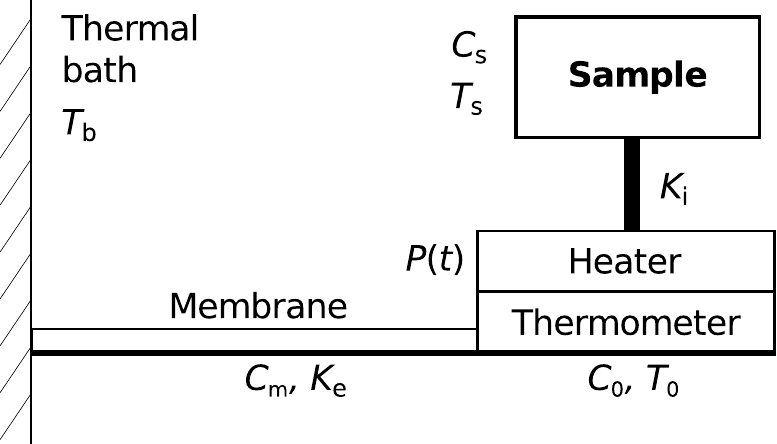}%
\end{center}
\caption{Thermal diagram of the studied system. Sample with heat capacity $C_\mathrm{s}$ and temperature $T_\mathrm{s}$ is coupled through a thermal conductance $K_\mathrm{i}$ to a platform ($C_\mathrm{0}$, $T_\mathrm{0}$) which, in turn, is connected to the thermal bath through the supporting membrane. Heater and thermometer are thin films lying on top of each other. They compose the platform together with the central part of the membrane. The heat capacity $C_\mathrm{0}$ represents an always existing background contribution of the calorimetric cell. The membrane outside the platform is treated either as a massless thermal conductance $K_\mathrm{e}$ or as a distributed object with a total heat capacity $C_{\mathrm{m}}$.}
\label{Fig_Scheme}
\end{figure}

In the primary work by Sullivan and Seidel~\cite{Sullivan} heater and thermometer were attached directly to the sample, which was connected to the thermal bath through a suitable thermal link. Such a design is not feasible in membrane-based calorimeters, where both thermometer and heater need to be thin films directly fabricated onto a free-standing membrane to maintain the heat capacity \hl{addenda} lower than the sample heat capacity. This separation of calorimeter cell from sample is not necessarily a drawback. It simplifies the system analysis and increases the experimental reproducibility, since there is only one thermal link to the sample that may vary from experiment to experiment rather than two or more. The most ideal situation is if, furthermore, both heater and thermometer are in good thermal contact with each other and with the membrane. The thermal diagram of such a system is depicted in Fig.~\ref{Fig_Scheme}. Heater, thermometer and membrane form a central sample platform, where the internal thermal coupling between thermometer and heater is assumed ideal. The platform is weakly connected, through the membrane thermal conductance $K_\mathrm{e}$, to a thermal bath at temperature $T_\mathrm{b}$. The sample is thermally connected to the platform by a minute amount of Apiezon grease or similar. A fundamental requirement for accurate measurements is that the thermal link $K_\mathrm{i}$ between sample and platform is much greater than $K_\mathrm{e}$.
The system of Fig.~\ref{Fig_Scheme} was analyzed by Velichkov~\cite{Velichkov} for the case corresponding to a massless thermal link. Riou~\textit{et al.}~\cite{RiouSM} extended the analysis to include the thermal diffusivity of sample and membrane, following the work by Greene~\textit{et al.}~\cite{Greene}. 

In this work we study the system both theoretically and experimentally. We show that the system model describes our experimental data well, provided that the frequency dependence of both the membrane \hl{addenda} and the membrane thermal link are taken into consideration for small samples. We finally suggest a criterion for the working frequency which ensures combined high resolution and good absolute accuracy.

\section{AC steady state measurements}
In the ac steady state method the temperature of sample and calorimetric cell is modulated with a small amplitude, giving rise to a temperature oscillation $T_\mathrm{ac}=T_\mathrm{ac,0}\sin(\omega t-\phi)$ of the thermometer.
The power responsible for this modulation in our case is due to resistive heating, and can be expressed as $P(t)=P_0(1+\sin\omega t)$, where $P_0=R_\mathrm{h}I^{2}_\mathrm{0}$, corresponding to an ac current with amplitude $I_\mathrm{0}\sqrt{2}$ and angular frequency $\omega'=\omega/2$ flowing through a resistor $R_\mathrm{h}$. With the use of a lock-in amplifier both the temperature oscillation amplitude $T_\mathrm{ac,0}$ and phase $\phi$ are experimentally accessible. 
They are given by \cite{Gmelin}
\begin{numcases}{}\label{EqTPhi}
\begin{aligned}
T_\mathrm{ac,0} &=\frac{P_\mathrm{0}}{\sqrt{(\omega C)^{2}+K^{2}}}	\\
\tan\phi &=\frac{\omega C}{K} 			
\end{aligned}
\end{numcases}
where, for the ideal case of a perfectly connected sample and massless membrane, $C=C_0+C_\mathrm{s}$ and $K=K_\mathrm{e}$. Equation~(\ref{EqTPhi}) can be reshaped to express the unknown $C$ and $K$ as a function of the measured parameters ($P_0$, $\omega$, $T_\mathrm{ac,0}$, $\phi$):
\begin{numcases}{}\label{EqCK}
\begin{aligned}
 C &=\frac{P_\mathrm{0}}{\omega T_\mathrm{ac,0}}\sin\phi		\\
 K &=\frac{P_\mathrm{0}}{T_\mathrm{ac,0}}\cos\phi 			
\end{aligned}
\end{numcases}
These expressions form the basics for evaluating ac steady state measurements.
At low frequency, $\omega \tau_\mathrm{e} \ll 1$, where
\begin{equation}
\tau_\mathrm{e}=(C_\mathrm{0}+C_\mathrm{s})/K_\mathrm{e}
\end{equation}
is the external time constant, the phase is close to 0 and the temperature response is dominated by the thermal link. In the opposite limit the phase ideally approaches 90$^{\circ}$. For $\omega \tau_\mathrm{e} > 7$ the absolute error arising from using the simple expression $C=P_\mathrm{0}/\omega T_\mathrm{ac,0}$ is less than 1\%. However, the conditions are rarely ideal at such frequencies. To obtain absolute accuracy, a good understanding of the system and a carefully selected frequency are therefore required.

When measuring small samples, $C_\mathrm{s} \lesssim C_\mathrm{m}$, the effect of the membrane should be considered. The thermal diffusion in the membrane introduces a new frequency dependence controlled by the parameter 
\begin{equation}
\tau_\mathrm{m}=C_\mathrm{m}/K_\mathrm{e}.
\end{equation}
For most samples the influence of the thermal link between sample and cell is also important. A weak connection between sample and platform introduces yet another time scale, described by the internal time constant
\begin{equation}
\tau_\mathrm{i}=C_\mathrm{s}/K_\mathrm{i}.
\end{equation}
We first study the effect of the membrane, assuming that the sample is well connected ($\tau_\mathrm{i}=0$). We then include the effect of a finite thermal link between sample and platform.

\section{Effect of membrane}
\begin{figure}[t]
\begin{center}
\includegraphics[width=0.75\linewidth]{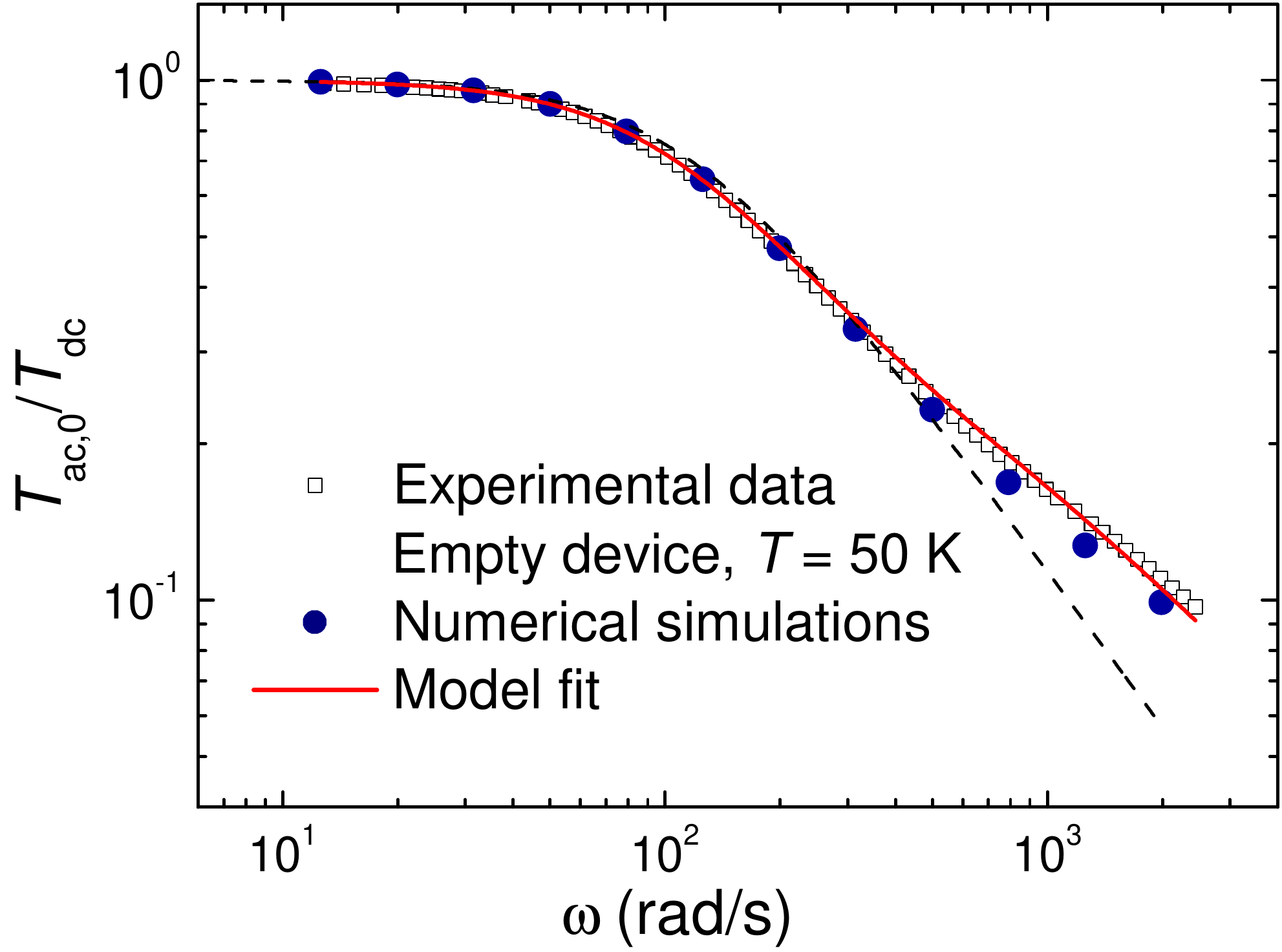}
\\[3mm]
\includegraphics[width=0.75\linewidth]{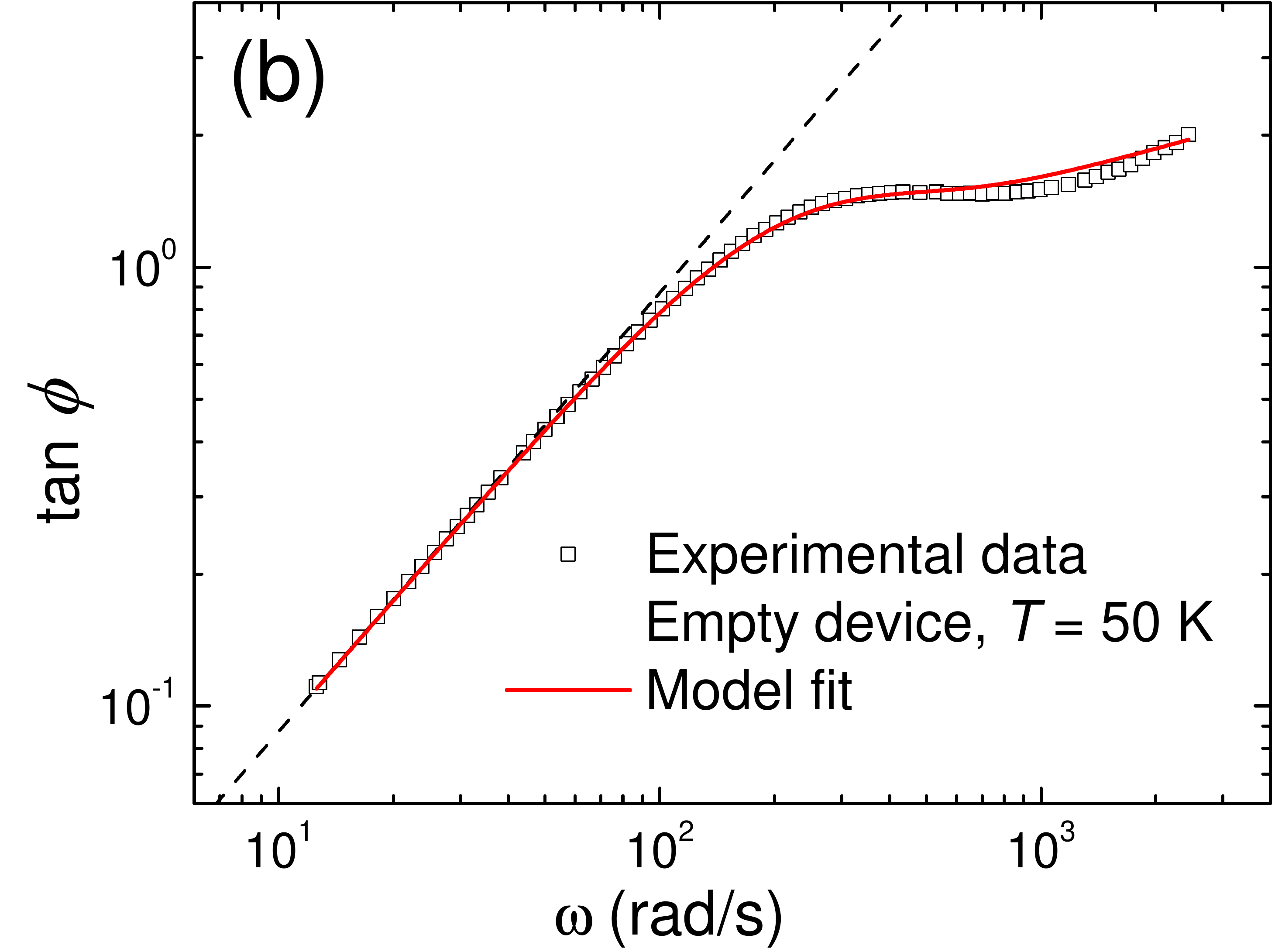}
\end{center}
\caption{Frequency dependence of $T_\mathrm{ac,0}$ and $\tan \phi$ for an empty cell ($C_\mathrm{s}=0$) at $T=50\,\mathrm{K}$. \textbf{a)} Temperature oscillation amplitude $T_\mathrm{ac,0}$ divided by its low frequency value $T_\mathrm{dc}=P_0/K_\mathrm{e}$. Numerical simulations are from \cite{Tagliati_Sim}. \textbf{b)} Phase angle $\phi$ between power and temperature, expressed as $\tan \phi$. The model fits are given by Eq.~(\ref{EqTPhi}) with $C=C_0+C_\mathrm{m,eff}$, $K=K_\mathrm{e,eff}$, and effective frequency dependences from Eq.~(\ref{EqCeff}) and (\ref{EqKeff}). The dashed curves were obtained by using the low-frequency limits $C=C_\mathrm{0}+C_\mathrm{m}/3$ and $K=K_\mathrm{e}$. The parameters in all cases are $C_\mathrm{0}=3.18\,\mathrm{nJ/K}$, $C_\mathrm{m}=33.2\,\mathrm{nJ/K}$, and $K_\mathrm{e}=1.63\,\upmu\mathrm{W/K}$.}
\label{Fig_Memb_TacPhi}
\end{figure}
\begin{figure}[t]
\begin{center}
\includegraphics[width=0.75\linewidth]{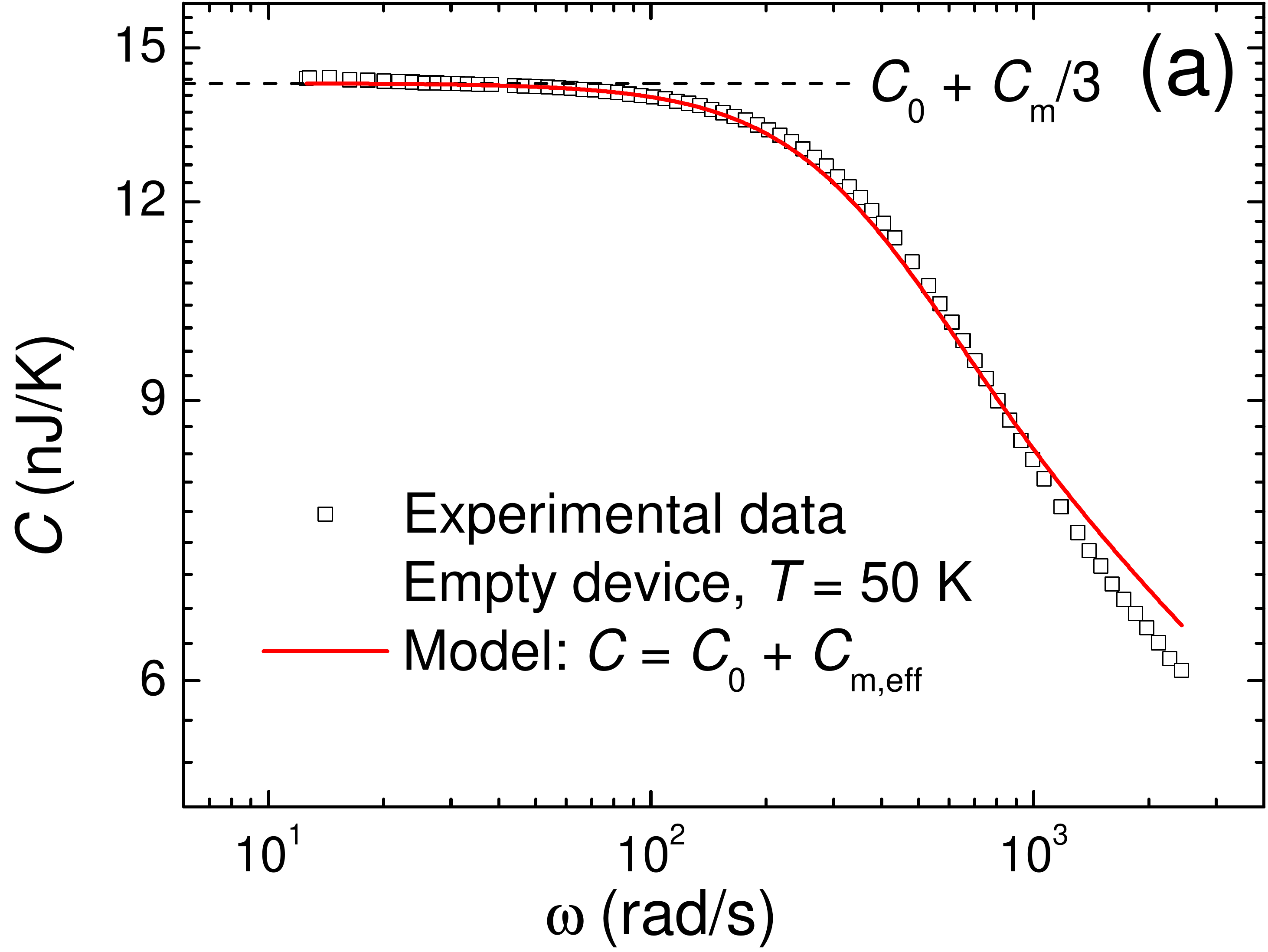}
\\[3mm]
\includegraphics[width=0.75\linewidth]{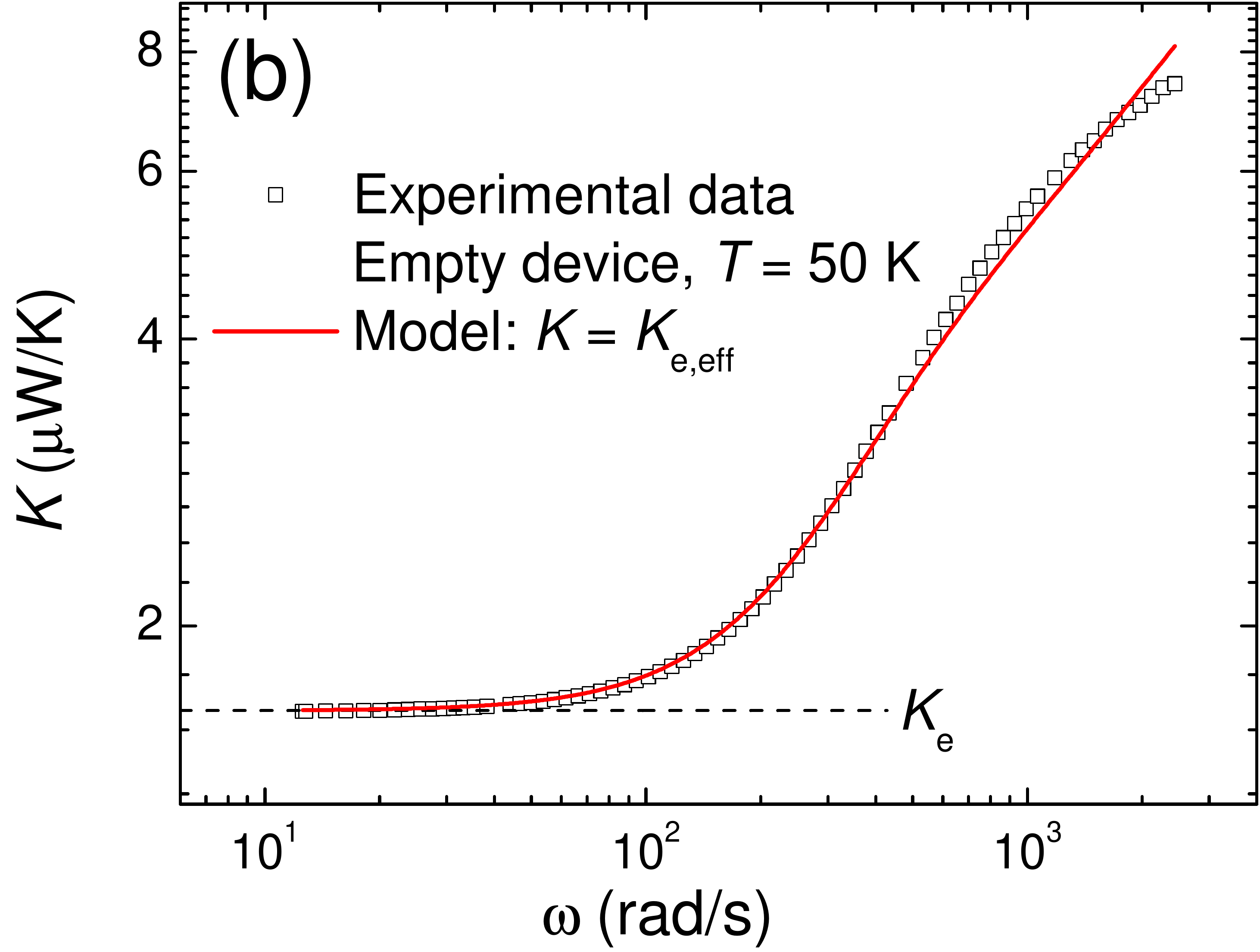}
\end{center}
\caption{Membrane properties. \textbf{a)} Effective membrane heat capacity. \textbf{b)} Effective thermal conductance. The experimental data were evaluated through Eq.~(\ref{EqCK}). All parameters, experimental data, and assumed frequency dependences are the same as in Fig.~\ref{Fig_Memb_TacPhi}.}
\label{Fig_Memb_CK}
\end{figure}
For small samples, the membrane can no longer be approximated as a massless thermal conductance, but should be seen as a total heat capacity $C_\mathrm{m}$ distributed over the membrane area. The generated temperature oscillation spreads in the membrane, including the metal leads of thermometer and heater, over a distance of the order of a frequency dependent thermal length $\ell_\mathrm{th}(\omega)=\sqrt{{2D}/{\omega}}$ \cite{RiouSM}, where $D=\kappa/\rho c_\mathrm{p}$ is the diffusivity, $\kappa$ the thermal conductivity, $\rho$ the density, and $c_\mathrm{p}$ the specific heat. For simplicity, we here model the membrane as a 1D rod. As shown in \ref{Freq 1D} for this model, following Greene \textit{et al.} \cite{Greene}, the diffusion practically results in an effective frequency dependent \hl{addenda} heat capacity from the membrane, given by
\begin{equation}
C_\mathrm{m,eff}=\frac{C_\mathrm{m}}{\alpha}\frac{\sinh\alpha-\sin\alpha}{\cosh\alpha-\cos\alpha},
\label{EqCeff}
\end{equation}
with $\alpha=\sqrt{2\omega \tau_\mathrm{m}}$. The limits for $C_\mathrm{m,eff}$ are $C_\mathrm{m}/3$ at low frequency and $C_\mathrm{m}/\alpha$ at high frequency. With increasing frequency the effective part that is temperature modulated thus shrinks, decreasing the membrane \hl{addenda} as $\omega^{1/2}$. 
However, not only $C=C_\mathrm{0}+C_\mathrm{s}+C_\mathrm{m,eff}$ becomes frequency dependent but also $K$. While the sensed heat capacity decreases, the thermal conductance increases because of a shorter effective distance of the temperature (amplitude) gradients. We show in \ref{Freq 1D} that it is natural to replace $K_\mathrm{e}$ by an effective thermal conductance
\begin{equation}
K_\mathrm{e,eff} = K_\mathrm{e} \frac{\alpha}{2}\frac{\sinh\alpha+\sin\alpha}{\cosh\alpha-\cos\alpha}.
\label{EqKeff}
\end{equation}
The low-frequency limit of $K_\mathrm{e,eff}$ is again $K_\mathrm{e}$, while the high-frequency limit is given by $\alpha K_\mathrm{e}/2$.

In Fig.~\ref{Fig_Memb_TacPhi} we apply Eq.~(\ref{EqCeff}) and (\ref{EqKeff}) to describe our experimental data of a calorimeter cell without sample. \hl{The calorimeter is made of thin film heater, electrical insulation and thermometer built on top of each other.  This stack covers the central $110\times110~\upmu\mathrm{m}^2$ area of a $1\times1~\mathrm{mm}^2$ and $150~\mathrm{nm}$ thick Si$_3$N$_4$ membrane, and form the platform onto which the sample is placed} \cite{Tagliati_Sim}.
The value of $K_\mathrm{e}$ is easily determined from $T_\mathrm{dc}$ at low frequency, while $C_\mathrm{m}$ is obtained from adjusting $\tau_\mathrm{m}$.
Figure~\ref{Fig_Memb_CK} shows the experimentally determined $C$ and $K$ for the empty cell, using Eq.~(\ref{EqCK}), with the corresponding model curves obtained by using the same parameter values as in Fig.~\ref{Fig_Memb_TacPhi}.
Note that $C_\mathrm{m,eff}$ is dominating over $C_\mathrm{0}$ at all frequencies. At the highest frequencies, $K$ seems to have a tendency to saturate and $C$ decreases slightly faster than expected. Whether this is due to an experimental problem, such as a phase distortion, or if a more realistic model would display a different high-frequency behavior remains an open question.

\section{Effect of thermal link to sample}
In the normal, experimental case, the effect of a non-zero internal time constant limits the upper frequency of measurements. The expressions for the stationary temperature oscillation amplitudes and phase shifts are derived in \ref{App2} for $C_\mathrm{m}=0$. Surprisingly, $T_\mathrm{ac,0}$ and $\tan\phi$ are still given by Eq.~(\ref{EqTPhi}) provided that $C$ and $K$ are defined by Eq.~(\ref{Eq_Cbar}) and Eq.~(\ref{Eq_Kbar}). Studying these equations, we see that in the low frequency limit ($\omega\tau_\mathrm{i}\ll1$), $C=C_0+C_\mathrm{s}$ and $K=K_\mathrm{e}$, i.e., the temperature is uniformly distributed in the sample-cell system. In the oppsite limit ($\omega\tau_\mathrm{i}\gg1$), the sample is thermally disconnected from the platform and just the heat capacity of the platform is probed. Nevertheless the presence of the sample is still sensed by an increase of the apparent thermal conductance, $K=K_\mathrm{e}+K_\mathrm{i}$.

From the solution (\ref{Eq_Solution}), we see that the actual sample temperature oscillation amplitude decreases faster than $T_\mathrm{ac,0}$ when the frequency increases,
\begin{equation} \label{Tac_sample}
T_\mathrm{ac,s}=T_\mathrm{ac,0}\sqrt{1-g},
\end{equation}
where $g$ is defined in Eq.~(\ref{g}), going from 0 to 1 with increasing frequency. The ac signal from a thermometer in direct contact with the sample consequently goes to zero for $\omega \tau_\mathrm{i} \gg 1$, while a thermometer on the cell, as in the present case, still gives useful information such as a rough estimate of the cell \hl{addenda}.

\begin{figure}[!ht]
\begin{center}
\includegraphics[width=0.75\linewidth]{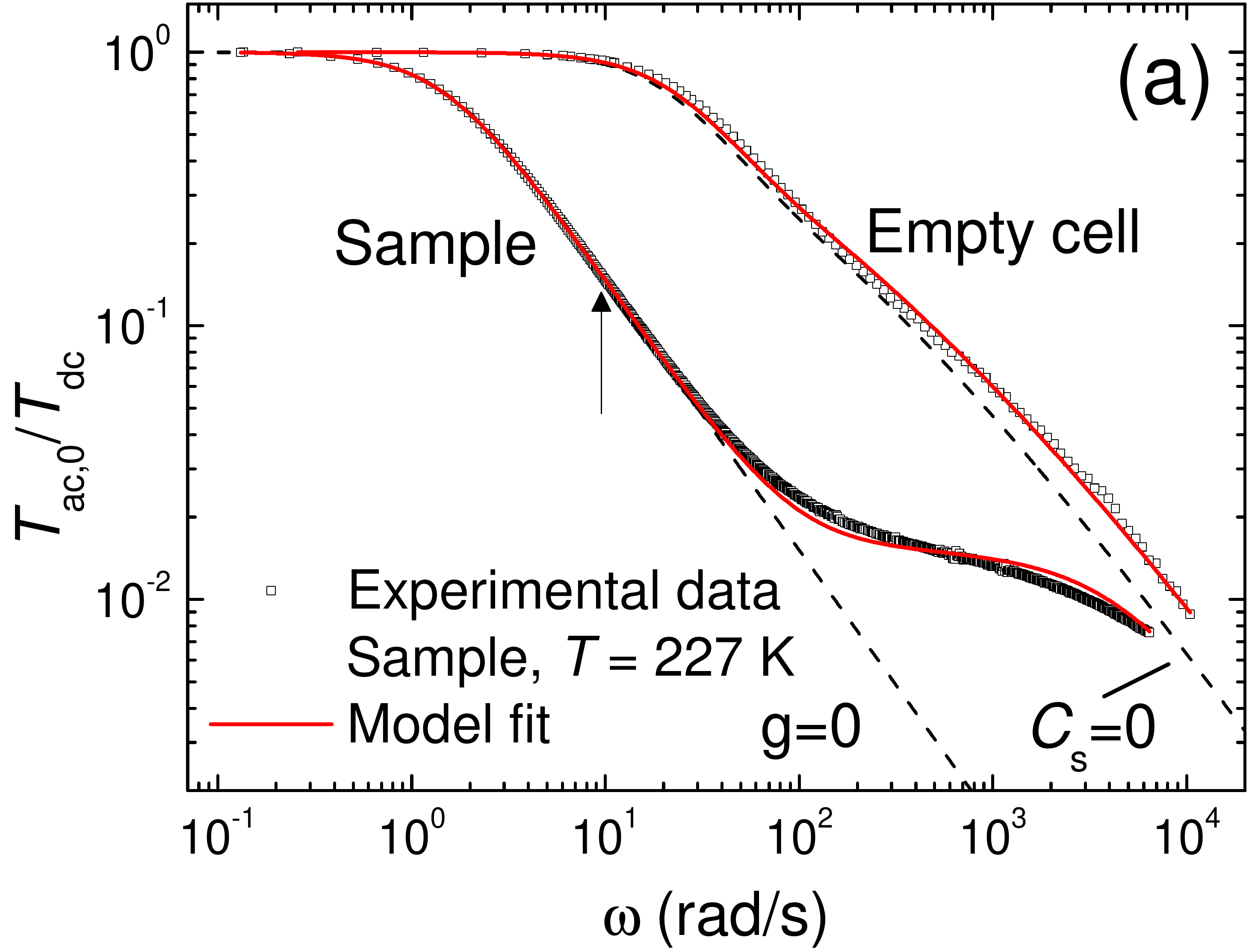}
\\[3mm]
\includegraphics[width=0.75\linewidth]{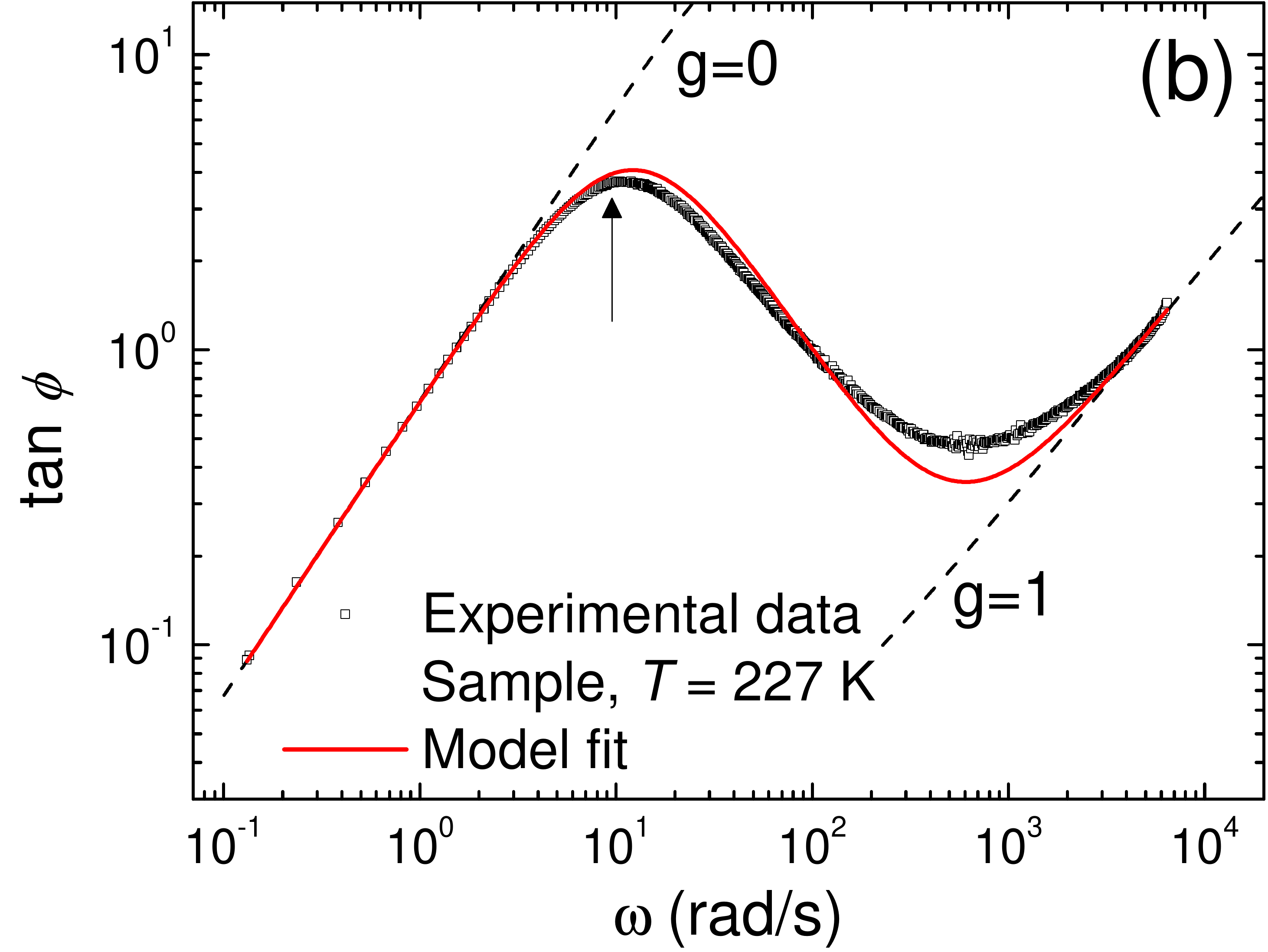}
\\[2mm]
\includegraphics[width=0.75\linewidth]{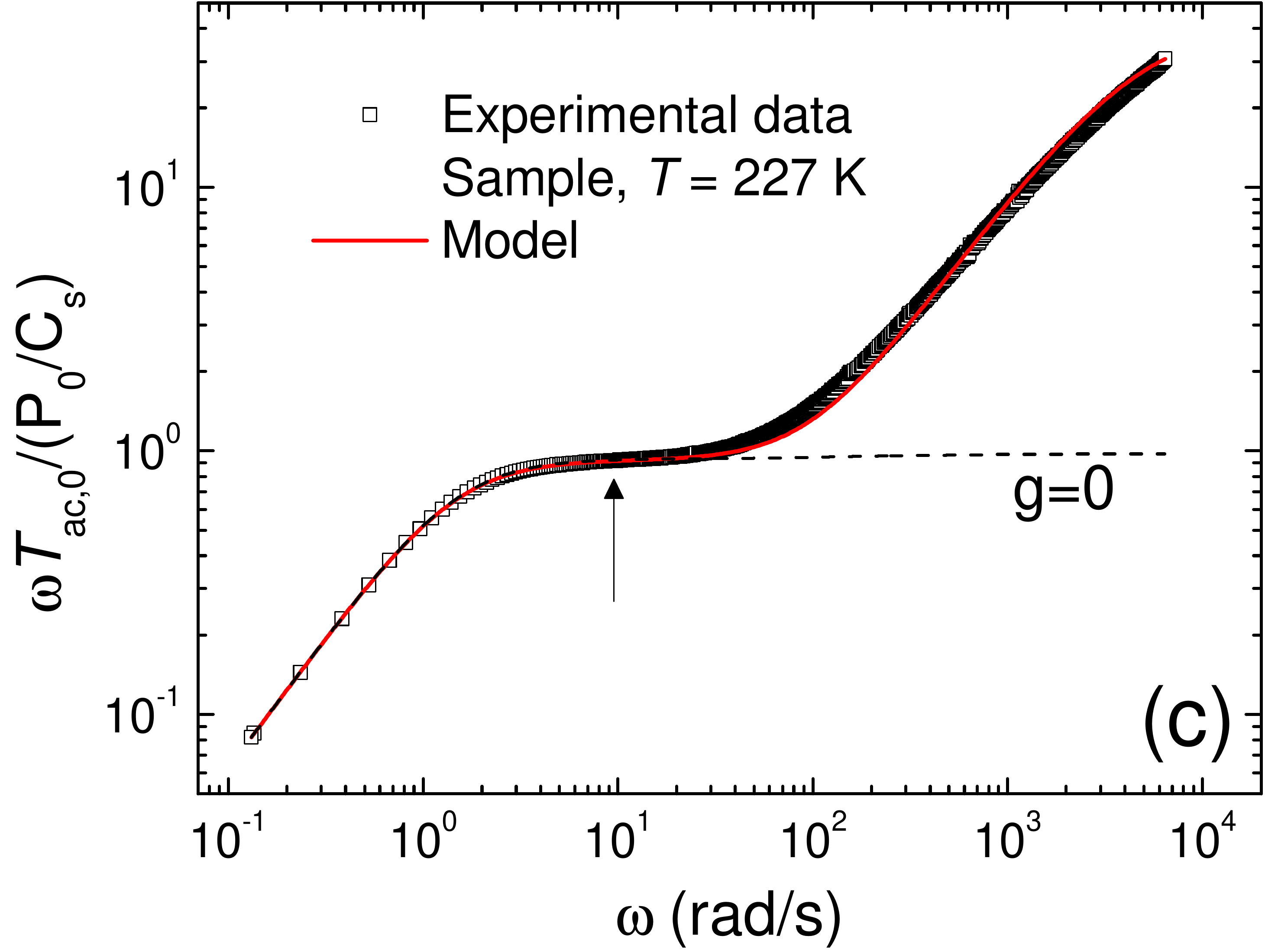}
\end{center}
\caption{\textbf{a)} Temperature oscillation amplitude of cell with sample and empty cell at $T=227\,\mathrm{K}$. Dashed curves correspond to a well-connected sample ($g=0$) and no sample ($C_\mathrm{s}=0$). \textbf{b)} $\tan \phi$ for cell with sample. Dashed curves correspond to a well-connected sample ($g=0$) and a disconnected sample ($g=1$). \hl{\textbf{c)} Transfer function of cell with sample.} The arrows indicate the frequency where 1\% of $C_\mathrm{s}$ has been decoupled ($g=0.01$). The fitting parameters of cell with sample are: $C_\mathrm{s}=1.93\,\upmu\mathrm{J/K}$, $K_\mathrm{e}=3.09\,\upmu\mathrm{W/K}$, $K_\mathrm{i}=187\,\upmu\mathrm{W/K}$, $C_\mathrm{m}=280\,\mathrm{nJ/K}$, and $C_\mathrm{0}=42\,\mathrm{nJ/K}$. The parameters for the empty cell measurement are $K_\mathrm{e}=3.00\,\upmu\mathrm{W/K}$ and $C_\mathrm{0}=25\,\mathrm{nJ/K}$.}
\label{Fig_Sample_TacPhi}
\end{figure}

Figure~\ref{Fig_Sample_TacPhi} shows $T_\mathrm{ac,0}$, $\tan \phi$ \hl{and the transfer function} in a typical measurement.
\begin{figure}[t]
\begin{center}
\includegraphics[width=0.75\linewidth]{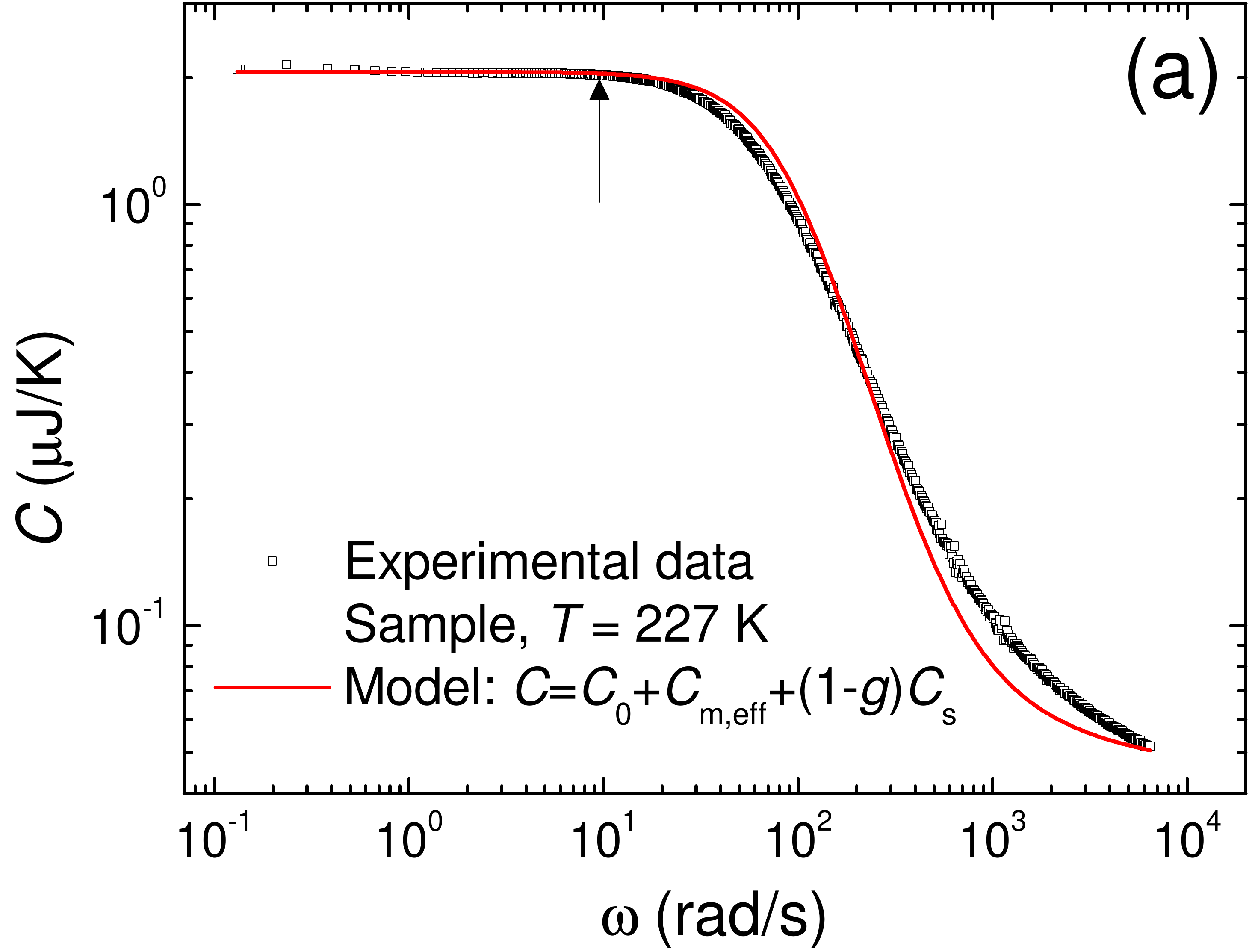}
\\[3mm]
\includegraphics[width=0.75\linewidth]{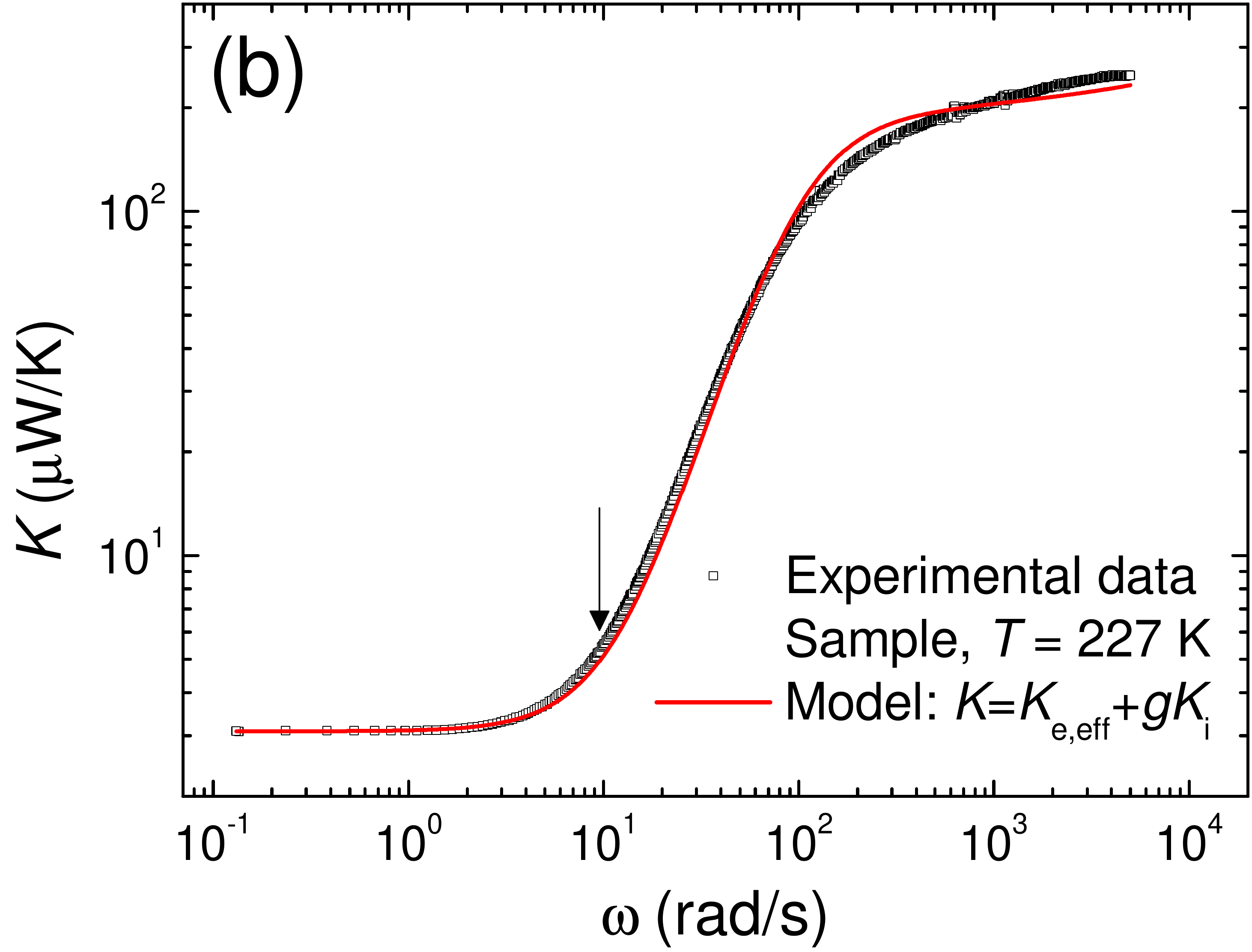}
\end{center}
\caption{\textbf{a)} Heat capacity of cell with sample. \textbf{b)} Thermal conductance. Model curves are given by Eq.~(\ref{Eq_CKtotal}) with the same parameter values as in Fig.~\ref{Fig_Sample_TacPhi}. The arrows indicate the frequency where 1\% of $C_\mathrm{s}$ has been decoupled ($g=0.01$).}
\label{Fig_Sample_CK}
\end{figure} 
The sample is a $\sim12.7\,\upmu\mathrm{g}$ piece of gold attached to the platform through Apiezon-N grease. A clear signature of the decoupling of the sample is seen both in $T_\mathrm{ac,0}$, which no longer goes as $1/\omega$, and in $\tan \phi$, which displays a characteristic decrease during the process. To fit the behavior of Fig.~\ref{Fig_Sample_TacPhi}, we introduce the effect of the membrane into the solution for a weakly connected sample by replacing $C_\mathrm{0}$ in Eq.~(\ref{Eq_Cbar}) with $C_\mathrm{0}+C_\mathrm{m,eff}$ and $K_\mathrm{e}$ in Eq.~(\ref{Eq_Kbar}) with $K_\mathrm{e,eff}$. We thus have
\begin{numcases}{}
\begin{aligned}
C &= C_0+C_\mathrm{m,eff} + (1-g)C_\mathrm{s}\\
K &=K_\mathrm{e,eff}+gK_\mathrm{i}
\end{aligned}\label{Eq_CKtotal}
\end{numcases}
This is a central result of the paper. The determination of the parameters is made easier by first fitting a measurement of $T_\mathrm{ac}$ for the empty cell, as shown in Fig.~\ref{Fig_Sample_TacPhi}a. This gives a well-determined value of $C_\mathrm{m}$. $K_\mathrm{e}$ is obtained from the known power and the low-frequency $T_\mathrm{ac,0}$, both for the empty cell case and for the case with sample. The values of $K_\mathrm{i}$, $C_\mathrm{s}$, and $C_\mathrm{0}$ are then obtained by fitting. A few observations can be pointed out. First, the values of $K_\mathrm{e}$ are slightly different for cell with sample and empty cell. This difference is within variations between experiments, depending on factors such as residual gas conduction and radiation. \hl{Second, the heat capacity for a Au piece with mass $m = 12.7\,\upmu\mathrm{g}$ at $T=227\,\mathrm{K}$ is about $1.59\,\upmu\mathrm{J/K}$ from literature. The difference between this value and the measured heat capacity is coming from the Apiezon grease. For good absolute accuracy, the heat capacity of the grease should thus be measured separately before attaching a sample.} Third, there is a difference between $C_\mathrm{0}$ for the empty membrane, and the residual $C_\mathrm{0}$ when the sample is disconnected. This difference can be ascribed to some Apiezon grease in good contact with the platform. Fourth, $C_\mathrm{0}$ of the empty membrane is in our case roughly 10\% of $C_\mathrm{m}$ at both discussed temperatures. The total membrane \hl{addenda} can therefore be estimated at any frequency provided that the temperature dependences of $C_\mathrm{m}$ and $K_\mathrm{e}$ are known. 

In Fig.~\ref{Fig_Sample_CK}, we show $C$ and $K$ as obtained from Eq.~(\ref{EqCK}) and fitted by Eq.~(\ref{Eq_CKtotal}), with the same parameters as in Fig.~\ref{Fig_Sample_TacPhi}.
In both Fig.~\ref{Fig_Sample_TacPhi} and \ref{Fig_Sample_CK}, arrows mark the frequency above which more than 1\% of the sample heat capacity signal has been lost ($g=0.01$ corresponding to $\omega \tau_\mathrm{i} \approx 0.1$). Note that $\tan\phi$ and $K$ are showing earlier signs of the decoupling than $T_\mathrm{ac,0}$ and $C$. Apparent $1/\omega$ behavior of $T_\mathrm{ac,0}$, \hl{which corresponds to the plateau of the transfer function in Fig.}~\ref{Fig_Sample_TacPhi}c, is thus not enough for good absolute accuracy. \hl{In the middle of the plateau, where experiments  are often performed, the absolute accuracy may already be worse than 1\%.}

\section{Optimizing the working frequency}
In practice, it would be too time consuming to measure the frequency dependence of $T_\mathrm{ac,0}$ and $\tan\phi$ at every temperature to determine $C_\mathrm{s}(T)$. It is therefore important to establish well-defined measurement conditions that ensure both good absolute accuracy and high resolution. This can be done by adjusting the frequency of the measurements, the working frequency $\omega_\mathrm{work}$. The best $\omega_\mathrm{work}$ depends on the requirements and limitations of the experiment: absolute accuracy, resolution, allowed dc offset $T_\mathrm{dc}$, and maximum amplitude of $T_\mathrm{ac,0}$. 

The resolution of the heat capacity is obtained by differentiating $C$ \hl{in Eq.}~(\ref{EqCK}) with respect to the temperature oscillation amplitude and phase, \hl{$\Delta C = \left|\partial C/\partial T_\mathrm{ac,0}\right|\Delta T_\mathrm{ac,0}+\left|\partial C/\partial \phi\right|\Delta \phi$}. For simplicity, we assume that the sample is well-connected and that the membrane frequency dependence can be neglected. We also assume that the phase resolution is given by $\Delta \phi=\Delta T_\mathrm{ac,0}/T_\mathrm{ac,0}$, where $\Delta T_\mathrm{ac,0}$ is determined by equipment and setup. Under these conditions
\begin{equation} \label{EqDeltaC}
\frac{\Delta C}{C}\approx\frac{\Delta T_\mathrm{ac,0}}{T_\mathrm{ac,0}}\left(1+\frac{1}{\tan\phi}\right).
\end{equation}
At constant $T_\mathrm{dc}$ (i.e.\ $P_0$), Eq.~(\ref{EqDeltaC}) is minimized for $T_\mathrm{ac,0}/T_\mathrm{dc}=\cos\phi=1/\sqrt{2}$ corresponding to $\phi= 45^\circ$ or $\tan\phi=1$. If we instead maintain a constant $T_\mathrm{ac,0}$, the resolution can be improved by a factor of 2 by increasing the frequency further, i.e., by decreasing the excess noise factor $1/\tan\phi$. 

Considering absolute accuracy, lower frequencies are generally more accurate as long as Eq.~(\ref{EqCK}) is used. Inaccuracies (beyond experimental problems etc.) are due to a nonzero $g$ in Eq.~(\ref{Eq_CKtotal}), in turn caused by a finite $K_\mathrm{i}$. As seen in Fig.~\ref{Fig_Sample_TacPhi}b, the sample starts to decouple near the local maximum of $\tan\phi$. The key parameter that controls this maximum is the ratio
\begin{equation}
\beta \equiv K_\mathrm{i}/K_\mathrm{e}.
\end{equation}
The location of the maximum is $\omega_{\max}\tau_\mathrm{e}\approx\sqrt{\beta}$, while its value is $\tan\phi_{\max}\approx\sqrt{\beta}/2$. At the maximum (still neglecting the membrane frequency dependence), the absolute error is
\begin{equation}
g_{\big|\tan \phi_{\max}}=\frac{1}{\beta+2} \approx \frac{1}{\beta}.
\end{equation}
At frequencies $\omega < \omega_{\max}$, where $\tan \phi \approx \omega\tau_\mathrm{e}$, the absolute error can be estimated to
\begin{equation}\label{EqAbsError}
g_{\big| \omega < \omega_{\max}} \approx {\left({\frac{\tan\phi}{\beta}}\right)}^2.
\end{equation}
With these rules of thumb, consider a few cases. To reach $\tan\phi =7$ ($\phi_{\max}\gtrsim 82^{\circ}$), as required for using the simplified relation $C=P_\mathrm{0}/\omega T_\mathrm{ac,0}$ with good absolute accuracy, a ratio $\beta \gtrsim 200$ is needed. To have less than 1\% error at the maximum of $\tan\phi$, $\beta$ must be greater than 100, corresponding to $\tan\phi_{\max} > 5$ ($\phi_{\max} \gtrsim 79^\circ$). Finally, to have less than 1\% error at $\tan\phi=1$, we only need $\beta > 10$, corresponding to $\tan\phi_{\max} > 1.58$ ($\phi_{\max} \gtrsim 58^\circ$).

From Eq.~(\ref{EqDeltaC}) and Eq.~(\ref{EqAbsError}) it is clear that the conditions are most well-controlled if $\omega$ is adjusted to maintain a constant phase $\phi_\mathrm{work}$. Higher $\tan\phi$ gives less noise, but around $\tan\phi_{\max}$ the accuracy decreases quickly and the system will become sensitive to changes in $\beta$. Looking at Eq.~(\ref{EqAbsError}), it would seem reasonable to choose $\tan\phi$ as a small fraction of $\beta$. However, such a criterion would quickly exceed $\tan\phi_{\max}$ for large $\beta$. A better choice is to take $\tan\phi$ as a fraction of $\tan\phi_{\max}$ which goes as $\sqrt{\beta}$. A suitable number is $\tan\phi= (2/3)\tan\phi_\mathrm{max}$, corresponding to
\begin{equation}
\tan\phi_\mathrm{work}\approx\sqrt{\beta}/3.
\end{equation}
This condition gives $g \approx 1/(9\beta)$ and thus less than $1\%$ error and $100\%$ excess noise as long as $\beta\gtrsim 10$. If we have $\beta=60$ as in Fig.~\ref{Fig_Sample_TacPhi}, we get $\tan\phi\approx 2.6$ ($\phi\approx 69^\circ$). The expected absolute error at this phase is $0.2\%$ and the excess noise is $38\%$, which is a fairly small number considering that a typical achievable resolution $\Delta C/C$ is  1 in $10^4$ to $10^5$. 

Since $\beta$ may vary during the measurements, it is important to be able to verifying that $g$ remains small. This can be done by studying $K$. At the point where $\tan\phi\approx\sqrt{\beta}/3$ the measured value of $K$ should be $\beta g \approx 10\%$ higher than $K_\mathrm{e}$, which in turn can be obtained from calibration measurements. In practice, $\beta$ is varying rather slowly, and $\phi_\mathrm{work}$ can often be maintained at a constant value.

\section{Conclusions}
The analysis carried out in this paper illustrates the care needed to obtain absolute accuracy in ac steady-state calorimeter measurements. Two problems that cannot be avoided when studying small samples are the frequency-dependent contribution of the sample support and the thermal link between sample and support. To handle these complications, it is not enough to present a model for the temperature oscillation and phase expressed as a function of time constants. Rather, explicit expressions for the sample heat capacity and external thermal link are needed as a function of experimentally determinable parameters.  Here we provide such expressions and show that they can be used to overcome the experimental obstacles. Based on the analysis, we argue that measurements are best performed at a constant phase $\phi$. With modern measurement electronics, such a condition is both feasible and practical.

\section*{Acknowledgments}
Financial support from the K.\ and A.\ Wallenberg foundation and the Swedish Research Council is gratefully acknowledged. We would like to thank V. M. Krasnov and Luca Argenti for useful discussions.

\appendix
\section{Membrane frequency dependence}
\label{Freq 1D}
We approximate the membrane and thin film leads by a uniform rod of length $L$ along the $x$-axis, connecting sample at $x=0$, at temperature $T_\mathrm{0}$, with thermal bath at $x=L$, at temperature $T_\mathrm{b}$. The profile of the temperature oscillation can be found from the heat equation
\begin{equation} \label{HeatEq}
\frac{\partial T}{\partial t}=D\frac{\partial^{2} T}{\partial x^{2}}.
\end{equation}
The boundary conditions of Eq.~(\ref{HeatEq}) are
\begin{align}
\label{EqBC1} &T(0,t)=T_\mathrm{0}+T_\mathrm{ac,0}\sin\omega t \\
\label{EqBC2} &T(L,t)=T_\mathrm{b} 
\end{align}
where the sample end is subjected to a power 
\begin{equation} \label{EqP}
P=P_\mathrm{dc}+P(t)=P_\mathrm{dc}+P_\mathrm{0}\sin(\omega t+\phi).
\end{equation}
Solving the time-independent problem gives the steady-state temperature profile:
\begin{equation} \label{TprofileDC}
T_\mathrm{dc}(x)=T_\mathrm{0}+\frac{x}{L}(T_\mathrm{b}-T_\mathrm{0}).
\end{equation}
The full solution is the sum of the time-independent and time-dependent contributions, $T(x,t)=T_\mathrm{dc}(x)+T_\mathrm{ac}(x,t)$. The steady-state temperature oscillation along the rod is found by means of Laplace transforms:
\begin{equation} \label{EqTprofile}
\frac{T_\mathrm{ac}(x,t)}{T_\mathrm{ac,0}}=\frac{\operatorname{Im}\left\{\sinh\left[\sqrt{\frac{-i\omega}{D}}(x-L)\right]\sinh\left(\sqrt{\frac{i\omega}{D}}L\right)e^{-i\omega t}\right\}}{\left|\sinh\left(\sqrt{\frac{i\omega}{D}}L\right)\right|^{2}}.
\end{equation}
This temperature oscillation is connected to the applied power through the Fourier law with an additional term for the sample heat capacity,
\begin{equation}\label{EqFourier}
P(t)=-LK_\mathrm{e}\frac{\mathrm{d}T_\mathrm{ac}(x,t)}{\mathrm{d}x}\Bigg|_{x=0}+C_\mathrm{s+0}\frac{\mathrm{d}T_\mathrm{ac}(x,t)}{\mathrm{d}t}\Bigg|_{x=0},
\end{equation}
where $K_\mathrm{e}$ is the thermal conductance of the rod and $C_\mathrm{s+0}$ is the sample heat capacity (including central platform). Combining Eqs.~(\ref{EqBC1}), (\ref{EqP}), (\ref{EqTprofile}), and (\ref{EqFourier}) gives
\begin{equation} \label{C_x0}
C_\mathrm{s+0}+\frac{C_\mathrm{m}}{\alpha}\frac{\sinh\alpha-\sin\alpha}{\cosh\alpha-\cos\alpha} = 
\frac{P_\mathrm{0}}{\omega T_\mathrm{ac,0}}{\sin\phi}\\
\end{equation}
and
\begin{equation} \label{K_x0}
K_\mathrm{e} \frac{\alpha}{2}\frac{\sinh\alpha+\sin\alpha}{\cosh\alpha-\cos\alpha} =
\frac{P_\mathrm{0}}{T_\mathrm{ac,0}}\cos\phi 
\end{equation}
where the diffusivity $D$ was substituted by $L^2K_\mathrm{e}/C_\mathrm{m}$ and the parameter $\alpha=\sqrt{2\omega\tau_\mathrm{m}}=2 L/\ell_\mathrm{th}$ with $\tau_\mathrm{m}=C_\mathrm{m}/K_\mathrm{e}$ was introduced. 
Equation~(\ref{C_x0}) and (\ref{K_x0}) take the shape of Eq.~(\ref{EqCK}) if we define the $\alpha$-containing terms on the left-hand side as the effective heat capacity and thermal link, respectively.

\section{Effect of a weakly connected sample}
\label{App2}
\noindent In the case of a weakly connected sample, one needs to keep track of the temperatures of both sample and platform. Below we make the following assumptions: (i) the temperature varies so little that the parameters are temperature independent, (ii) the thermal links are massless, and (iii) the thermal conductances of thermometer, heater and sample are infinite (as before).

The thermal response of the platform and sample can be written as
\begin{subequations}
\begin{align}
T_\mathrm{0}(t)=T_\mathrm{b}+T_\mathrm{off}+T_\mathrm{ac,0}(t) \\
T_\mathrm{s}(t)=T_\mathrm{b}+T_\mathrm{off}+T_\mathrm{ac,s}(t)
\end{align}
\end{subequations}
where $T_\mathrm{off}$ is the dc offset due to the time-averaged power supplied by the heater resistance and any other dc power. $T_\mathrm{ac}(t)$ is the oscillating term that can be further expressed as
\begin{subequations} \label{Tmod}
\begin{align}
T_\mathrm{ac,0}(t)=T_\mathrm{ac,0}\sin(\omega t-\phi) \\
T_\mathrm{ac,s}(t)=T_\mathrm{ac,s}\sin(\omega t-\varphi)
\end{align}
\end{subequations}
where $\phi$ and $\varphi$ are the phase shifts which develop between power $P (t)=P_\mathrm{dc}+P_{0}\sin\omega t$ and temperature $T_\mathrm{ac}(t)$, due to the finite thermal conductances $K_\mathrm{e}$ and $K_\mathrm{i}$. 
Expressions for the measured variables can be obtained with some effort. The thermal equations governing the system ensure the conservation of energy for each part of the system:
\begin{subequations}
\begin{align} 
C_\mathrm{0}\frac{\mathrm{d}T_\mathrm{0}}{\mathrm{d}t}&=P+K_\mathrm{e}(T_\mathrm{b}-T_\mathrm{0})+K_\mathrm{i}(T_\mathrm{s}-T_\mathrm{0}), \label{BalanceEqR}\\
C_\mathrm{s}\frac{\mathrm{d}T_\mathrm{s}}{\mathrm{d}t}&=K_\mathrm{i}(T_\mathrm{0}-T_\mathrm{s}) \label{BalanceEqS}.
\end{align}
\end{subequations}
Inserting the full expressions for $P$, $T_\mathrm{s}$ and $T_\mathrm{0}$ in Eqs.~(\ref{BalanceEqR}) and (\ref{BalanceEqS}), we get one time-independent equation describing the temperature offset
\begin{equation}
T_\mathrm{off} = \frac{P_\mathrm{dc}}{K_\mathrm{e}},\label{ConstEq}\\
\end{equation}
and two time-dependent equations,
\begin{numcases}{}
\begin{aligned}
\left[\frac{\omega C_\mathrm{0}}{\tan(\omega t-\phi)}+K_\mathrm{e}+K_\mathrm{i} \right]X_0=& K_\mathrm{i}X_s + P_\mathrm{0}\sin\omega t \label{TimeDepEq1}\\
\left[\frac{\omega C_\mathrm{s}}{\tan(\omega t-\varphi)}+K_\mathrm{i} \right]X_s=& K_\mathrm{i}X_0 \label{TimeDepEq2}
\end{aligned}
\end{numcases}
where
$X_0 = T_\mathrm{ac,0}\sin(\omega t-\phi)$ and $X_s = T_\mathrm{ac,s}\sin(\omega t-\varphi)$. Equation~(\ref{TimeDepEq1}) can be expanded through some trigonometric identities before the terms multiplying $\sin\omega t$ and $\cos\omega t$ are collected. This results in a system of four unknowns:
\begin{equation}
 \begin{pmatrix}
  \omega C_\mathrm{0} & K_\mathrm{e}+K_\mathrm{i} & 0 & -K_\mathrm{i} \\
  K_\mathrm{e}+K_\mathrm{i} &  -\omega C_\mathrm{0} & -K_\mathrm{i} & 0 \\
  0  & -K_\mathrm{i}  & \omega C_\mathrm{s} & K_\mathrm{i}  \\
  K_\mathrm{i} & 0 & -K_\mathrm{i}  & \omega C_\mathrm{s} 
 \end{pmatrix}
  \begin{pmatrix}
 X_1 \\
  X_2 \\
  X_3  \\
  X_4
 \end{pmatrix}
 =
  \begin{pmatrix}
 P_\mathrm{0} \\
  0 \\
  0  \\
  0
 \end{pmatrix}
\end{equation}
where
\begin{equation}
 \begin{pmatrix}
 X_1 \\
  X_2 \\
  X_3  \\
  X_4
 \end{pmatrix}
 =
  \begin{pmatrix}
T_\mathrm{ac,0}\sin\phi \\
T_\mathrm{ac,0}\cos\phi \\
T_\mathrm{ac,s}\sin\varphi \\
T_\mathrm{ac,s}\cos\varphi
 \end{pmatrix}
\end{equation}
The solution to this equation system is
\begin{equation}\label{Eq_Solution}
 \begin{pmatrix}
 X_1 \\
  X_2 \\
  X_3  \\
  X_4
 \end{pmatrix}
 = A
  \begin{pmatrix}
\omega C \\
K  \\
(\omega C+K \omega \tau_\mathrm{i})({1-g}) \\
(K-\omega C\omega \tau_\mathrm{i})({1-g})
 \end{pmatrix}
\end{equation}
where
\begin{equation} \label{Eq_A}
A= \dfrac{P_\mathrm{0}}{(\omega C)^2+K^2},
\end{equation}
\begin{equation} \label{Eq_Cbar}
C=C_0+(1-g)C_\mathrm{s},
\end{equation}
\begin{equation} \label{Eq_Kbar}
K=K_\mathrm{e}+gK_\mathrm{i}
 \end{equation}
and
\begin{equation} \label{g}
g=\frac{(\omega \tau_\mathrm{i})^2}{1+(\omega \tau_\mathrm{i})^2}.
\end{equation}
The solution given in this form can be easily reshaped into Eq.~(\ref{EqTPhi}) for the experimentally measured values $T_\mathrm{ac,0}$ and $\phi$ if $C$ and $K$ are defined as above.





\bibliographystyle{model1-num-names}
\bibliography{<your-bib-database>}

\begin{thebibliography}{00}


\bibitem{Sullivan}
P.~F.~Sullivan, G.~Seidel, Phys.~Rev. 173, (1968) 679.

\bibitem{RiouRSI}
O.~Riou, P.~Gandit, M.~Charalambous, J.~Chaussy, Rev.~Sci.~Instrum. 68 (1997) 1501.

\bibitem{Minakov}
A.~A.~Minakov, S.~B.~Roy, Y.~V.~Bugoslavsky, L.~F.~Cohen, Rev.~Sci.~Instrum. 76 (2005) 043906.

\bibitem{Huth}
H.~Huth, A.~A.~Minakov, C.~Schick, J.~Polym.~Sci.~B 44 (2006) 2996.

\bibitem{RydhEMSAT}
A.~Rydh, in Encyclopedia of Materials: Science and Technology, Online Update. K.~H.~J.~ Buschow, M.~C.~Flemings, R.~W.~Cahn, P.~Veyssi\`{e}re, E.~J.~Kramer and S.~Mahajan (eds.) Elsevier Ltd., Oxford (2006).

\bibitem{Garden}
J.-L.~Garden., H.~Guillou, A.~F.~Lopeandia, J.~Richard, J.-S.~Heron, G.~M.~Souche, F.~R.~Ong, B.~Vianay, O.~Bourgeois, Thermochim.~Acta 492 (2009) 16.

\bibitem{TagliatiCalori}
S.~Tagliati, A.~Rydh, R.~Xie, U.~Welp, W.~K.~Kwok, J.~Phys.:~Conf.~Ser. 150 (2009) 052256.

\bibitem{Kohama}
Y.~Kohama, C.~Marcenat, T.~Klein, M.~Jaime, Rev.~Sci.~Instrum. 81 (2010) 104902.

\bibitem{RiouSM}
O.~Riou, J.~F. Durastanti, Y.~Sfaxi, Superlattices and Microstructure 35 (2004) 353.

\bibitem{Velichkov}
I.~V.~Velichkov, Cryogenics 32 (1992) 285.

\bibitem{Greene}
R.~L.~Greene, C.~N.~King, R.~B.~Zubeck, J.~J.~Hauser, Phys.~Rev.~B 6 (1972) 3297.

\bibitem{Gmelin}
E.~Gmelin, Thermochim.~Acta 29 (1997) 1.

\bibitem{Tagliati_Sim}
S.~Tagliati, J.~Pipping, A.~Rydh, J.~Phys.:~Conf.~Ser. 234 (2010) 042036.

\end{thebibliography}



\end{document}